\documentclass[aps,twocolumn,english,superscriptaddress,citeautoscript,preprintnumbers,amsmath,amssymb,floatfix,footinbib]{revtex4-1}
\usepackage{hyperref}
\hypersetup{colorlinks=true,linkcolor=red,citecolor=blue}
\usepackage{amsmath}
\usepackage{graphicx}
\usepackage{dcolumn}
\usepackage{float}

\begin{document}

\title{Signature of topological non-trivial band structure in Ta$_{3}$SiTe$_{6}$}

\author{Shubhankar Roy}

\affiliation{Saha Institute of Nuclear Physics, HBNI, 1/AF Bidhannagar, Kolkata 700 064, India}

\affiliation{Vidyasagar Metropolitan College, 39, Sankar Ghosh Lane, Kolkata 700 006, India}

\author{Ratnadwip Singha}

\thanks{Present address: Department of Chemistry, Princeton University, Princeton, New Jersey 08544, USA}

\affiliation{Saha Institute of Nuclear Physics, HBNI, 1/AF Bidhannagar, Kolkata 700 064, India}

\author{Arup Ghosh}

\affiliation{Saha Institute of Nuclear Physics, HBNI, 1/AF Bidhannagar, Kolkata 700 064, India}

\author{Prabhat Mandal}

\thanks{prabhat.mandal@saha.ac.in}

\affiliation{Saha Institute of Nuclear Physics, HBNI, 1/AF Bidhannagar, Kolkata 700 064, India}

\date{\today}

\begin{abstract}

The study of topology protected electronic properties is a fascinating topic in present day condensed matter physics research. New topological materials are frequently being proposed and explored through various experimental techniques. Ta$_{3}$SiTe$_{6}$ is a newly predicted topological semimetal with fourfold degenerate nodal-line crossing in absence of spin-orbit coupling (SOC) and an hourglass Dirac loop, when SOC is included. Recent angle-resolved photoemission spectroscopy study in this material, has also confirmed Dirac like dispersions and two nodal-lines near the Fermi energy, protected by nonsymmorphic glide mirror symmetry. In this work, we present the detailed magnetotransport properties of single crystalline Ta$_{3}$SiTe$_{6}$. A nonsaturating magnetoresistance has been observed. Hall measurements reveal hole type charge carriers with high carrier density and a moderate value of carrier mobility. Furthermore, we report a robust planar Hall effect, which persists up to high temperatures. These results validate the nontrivial nature of the electronic band structure.

\end{abstract}

\maketitle

\section{Introduction}

In last few years, the nontrivial topological phases of matter have introduced a new paradigm of understanding the solids in condensed matter physics research. The discoveries of topological insulators (TI) \cite{Xia,Zhang,Chen}, Dirac and Weyl semimetals \cite{Liu,Liu2,Lv,Xu}, nodal-line materials \cite{Bian,Singha}, and topological superconductors \cite{Xu2,Matano}, have paved the way to realize and understand the dynamics of the relativistic quasiparticles such as Dirac, Weyl, and Majorana fermions in low-energy electronic systems. Topological materials, which host nodal-line fermions, are recently receiving considerable attention due to their various interesting properties like anisotropic electron transport \cite{Mullen}, possible surface magnetism/superconductivity \cite{Bian,Heikkila}, anomalous Landau level spectrum, etc \cite{Rhim,Lim}. Nodal-line materials with nonsymmorphic symmetry are of greater interest as they are robust against spin-orbit coupling (SOC), which often gaps out band crossing points. Moreover, nonsymmorphic symmetry protected band structure are expected to show more exotic quantum phases such as M$\ddot{o}$bius-twist surface states, hourglass fermions, and nodal chains, etc. \cite{Bzdusek,Wang4,Wang5,Ma,Shiozaki}.

Recently, from the first-principles calculations, $X_{3}$SiTe$_{6}$ ($X$=Ta, Nb) are predicted to show rich band crossing features with nodal fermions protected by nonsymmorphic space-group symmetry. In the absence of SOC, these materials host an essential fourfold nodal-line and an accidental nodal-loop \cite{Li6}. Furthermore, when the SOC is included, an hourglass Dirac-loop appears in the close vicinity of E$_{F}$. Even when the materials are thinned down to monolayers, the nontrivial band structure is sustained; predicted to have either a two-dimensional (2D) nodal-lines or 2D Dirac points in the absence or presence of SOC, respectively. An angle-resolved photoemission spectroscopy (ARPES) study on the layered ternary telluride Ta$_{3}$SiTe$_{6}$ has reported Dirac-like dispersions and the existence of two nodal-lines near the Fermi energy protected by the nonsymmorphic glide mirror symmetry of the crystal \cite{Sato}. On the other hand, from recent quantum oscillation measurements, it was not possible to confirm the non-trivial Berry phase due to the comparatively high oscillatory frequency and the presence of multiple hole pockets \cite{Naveed}. A candidate material with nodal-line band crossings near Fermi energy enabled by nonsymmorphic symmetry is still rare. Therefore, materials like $X_{3}$SiTe$_{6}$ with such intriguing predictions are highly desirable as they provide a fantastic platform to explore the exotic topological properties via various theoretical and experimental tools.

Sophisticated techniques like ARPES are frequently used as a direct experimental probe to reveal the topologically non-trivial band structure in materials. Whereas, albeit indirectly, by tailoring the temperature at a very low value and modulating the magnetic field, some transport measurements can also reveal non-trivial electronic structure close to the Fermi energy. In recent times, unusual experimental observations, e.g., large and non-saturating positive magnetoresistance (MR), ultrahigh carrier mobility, and low carrier effective mass, have been treated as the general characteristics of topological materials \cite{Singha,Ali,Shekhar,Roy}. The transport phenomena not only provide the signature of relativistic excitations but also evaluate the viability of the material for technological applications. Negative longitudinal MR (LMR) driven by Adler-Bell-Jackiw (ABJ) chiral anomaly under parallel electric and magnetic fields, is commonly regarded as definitive evidence of Dirac/Weyl fermions \cite{Huang,Li,Li2,Pariari,Wang}. However, because of the large positive MR at transverse electric and magnetic field configuration, a small misalignment between them can easily mask the weak negative LMR component \cite{Huang}. Furthermore, it has been reported that current jetting \cite{Hu} and weak localization effect \cite{Ulmet} can also produce negative LMR. Recently a new phenomenon, called the planar Hall effect (PHE) has been proposed to be directly related to the ABJ chiral anomaly and nontrivial Berry curvature \cite{Burkov,Nandy}. In PHE, the Hall voltage terminals, electric and magnetic fields are coplanar, whereas they are mutually perpendicular to each other in a conventional Hall measurement. As PHE is entirely distinct from conventional Hall effect, both in experimental configuration and angle dependence, it is easier to identify and confirm the topological non-trivial band structure in a material. PHE has already been observed in topological insulator ZrTe$_{5}$ \cite{Li3}; Dirac semimetals Cd$_{3}$As$_{2}$ \cite{Li4} and VAl$_{3}$ \cite{Singha3}; Weyl semimetals WTe$_{2}$ \cite{Wang2} and GdPtBi \cite{Kumar}. Although conventional metals and semiconductors are not expected to show PHE, a weak PHE can be observed in ferromagnetic compounds. However, for ferromagnets, it appears due to anisotropic MR, which originates from different in-plane and out-of-plane spin scattering \cite{Nazmul}.

In this paper, we have grown single crystal of proposed topological semimetal Ta$_{3}$SiTe$_{6}$ and measured the magnetotransport properties. Nonsaturating MR has been observed. The positive and linear field dependence of Hall resistivity confirms the presence hole type carriers. A prominent PHE has been observed, which is detectable up to high temperature and reveals the nontrivial nature of the band structure in this material.

\section{Experimental details}

Single crystals of Ta$_{3}$SiTe$_{6}$ were grown by standard chemical vapor transport technique. Iodine was used as a transport medium. The stoichiometric mixture of high-purity Ta powder (Alfa aesar 99.9\%), Si pieces (Alfa aesar 99.9999\%) and Te pieces (Alfa aesar 99.999\%) along with a small amount of iodine (Alfa aesar 99.5\%) was sealed in a quartz tube under high vacuum. The quartz tube was then placed in a two-zone horizontal furnace. The source end of the quartz tube containing the mixture and iodine was heated to 950$^{\circ}$C, while the other end was kept at 850$^{\circ}$C. This temperature difference was maintained for 7 days. The furnace was then cooled very slowly to room temperature. Several shiny, plate-like crystals formed at the cold end of the tube were mechanically extracted for transport measurements. Phase purity and the structural analysis on crashed single crystals of Ta$_{3}$SiTe$_{6}$ were done using the powder x-ray diffraction (XRD) technique with Cu-K$_{\alpha}$ radiation in a high resolution Rigaku x-ray diffractometer (TTRAX III). Within the resolution of XRD, we did not see any peak due to the impurity phase (Fig. 1). From Rietveld profile refinement, we have extracted the lattice parameters $a$=6.328 {\AA}, $b$=11.412 {\AA}, $c$=14.017 {\AA} with space group symmetry $Pnma$. The elemental composition was checked by energy dispersive x-ray (EDX) spectroscopy in a field emission scanning electron microscope [FESEM SUPRA 35 VP, Carl Zeiss (Germany)]. Mapping of the elemental concentrations at different randomly selected regions of the grown crystals provides detailed information about the overall chemical composition and confirms almost perfect stoichiometry [Ta:Si:Te=3:1.1:6.05] of the Ta$_{3}$SiTe$_{6}$ crystals. The relative error in the calculated atomic ratio is about 5 \%. The transport measurements were performed using standard four-probe technique in a 9 T physical property measurement system (Quantum Design) using the ac-transport option and sample rotator. Freshly cleaved crystals were used for transport measurements. Electrical contacts were made using gold wire and conducting silver paste.

\begin{figure}
\includegraphics[width=0.5\textwidth]{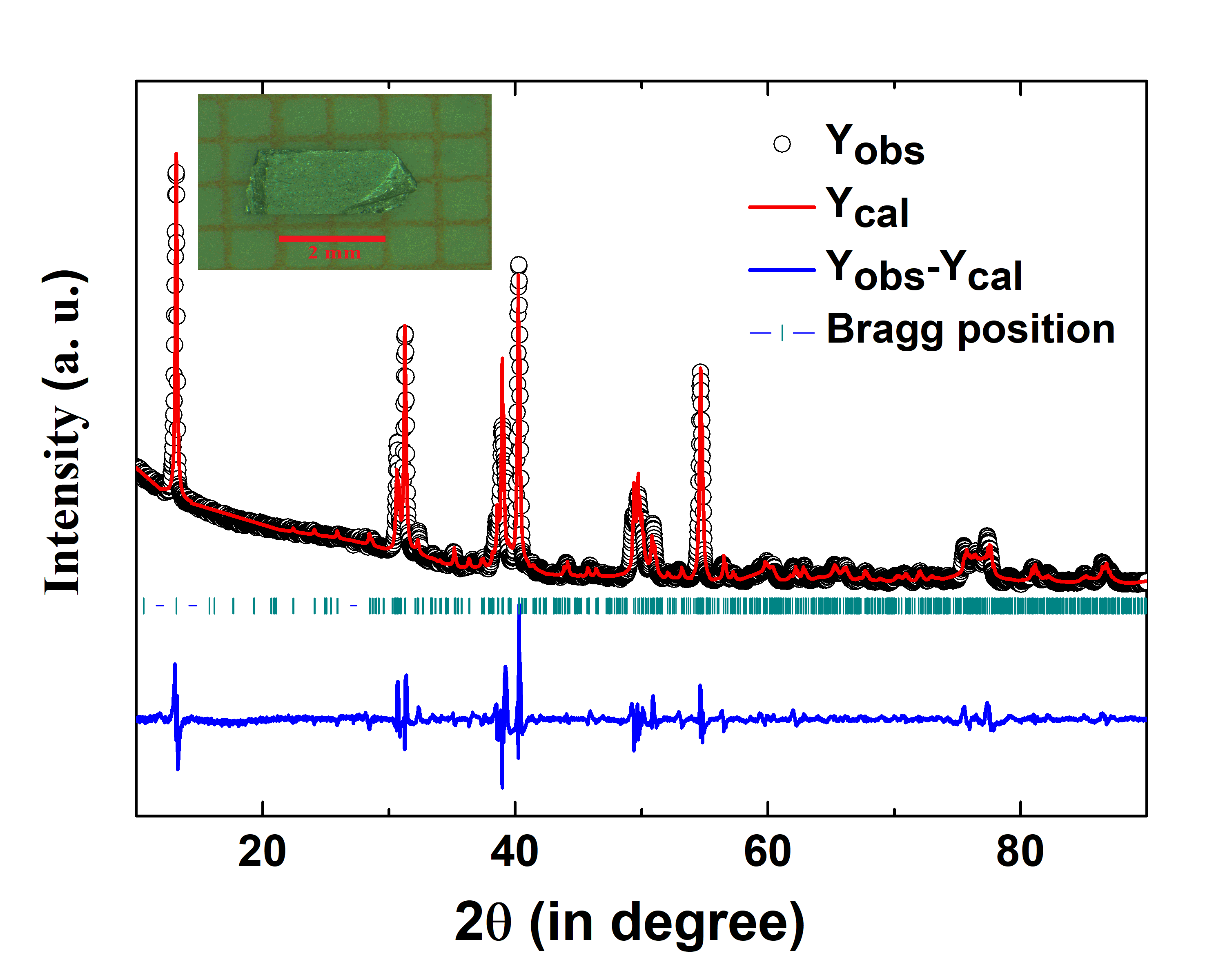}
\caption{ (a) Rietveld profile refinement of the powder XRD pattern for Ta$_{3}$SiTe$_{6}$. Black circles are experimental data (Y$_{obs}$), red line is the calculated pattern (Y$_{cal}$), blue line is the difference between experimental and calculated intensities (Y$_{obs}$-Y$_{cal}$), and magenta vertical lines are the Bragg positions. Inset shows a typical single crystal with scale bar for reference.}\label{rh}
\end{figure}

\begin{figure}
\includegraphics[width=0.5\textwidth]{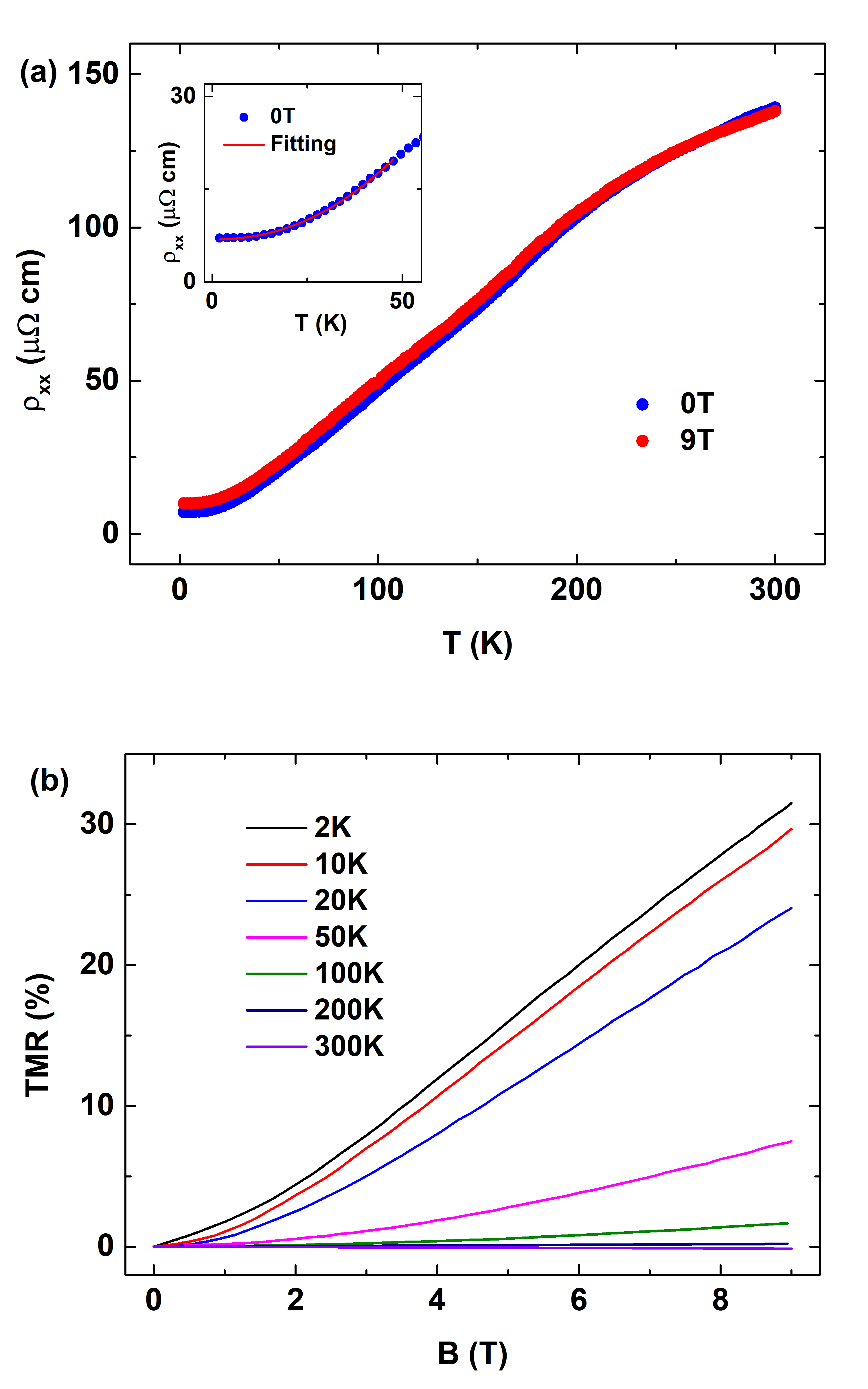}
\caption{ (a) Temperature dependence of resistivity ($\rho_{xx}$) both in presence and absence of transverse magnetic field. Inset shows the low-temperature region, fitted with $\rho_{xx}(T)=a+bT^2$. (b) Magnetic field dependence of transverse magnetoresistance of Ta$_{3}$SiTe$_{6}$ at different temperatures.}\label{rh}
\end{figure}

\section{Results and Discussions}

In Fig. 2(a), the longitudinal resistivity ($\rho_{xx}$) of a representative Ta$_{3}$SiTe$_{6}$ single crystal is plotted as a function of temperature, which shows metallic behavior down to lowest measured temperature. At room temperature, $\rho_{xx}$ is 139 $\mu$$\Omega$ cm. With decrease in temperature, $\rho_{xx}$ deceases monotonically and becomes 6.8 $\mu$$\Omega$ cm at 2 K. The residual resistivity ratio (RRR), $\rho_{xx}$(300 K)/$\rho_{xx}$(2 K), is $\sim$20, which indicates the good metallicity of the single crystal. As temperature decreases below 50 K, $\rho_{xx}$($T$) curve exhibits a weak upward curvature. This indicates a crossover in charge scattering mechanism with decrease in temperature. As shown in the inset, the zero-field resistivity at low temperature below 50 K can be fitted well with the equation, $\rho_{xx}$($T$)=$\rho_0$+$aT^{2}$. The quadratic temperature dependence implies pure electronic correlation dominated scattering in the low-temperature region \cite{Ziman}. The coefficient $a$ determines the strength of electron-electron scattering in the system. The low-temperature fit reveals $a$$\sim$ 6$\times$10$^{-3}$ $\mu$$\Omega$ K$^{-2}$. We note that usually, the zero-field resistivity in topological semimetals shows quite different temperature dependence. $\rho_{xx}$ is either very weakly dependent on $T$ or follows the relation $\rho_{xx}(T)\propto T^{n}$ with $n\geq$3 \cite{Singha,Shekhar,Tafti,Singha4}. Nonetheless, $T^{2}$ dependence of $\rho_{xx}$ has been reported in few topological semimetals \cite{Roy,Wang3}. Often the low-temperature resistivity of topological semimetals is strongly influenced by magnetic field. When the magnetic field is applied perpendicular to the current direction, the resistivity is observed to increase at low temperature [Fig. 2(a)]. However, there is no field-induced metal to semiconductor-like crossover as observed in most topological semimetals \cite{Singha,Ali,Shekhar,Roy,Pariari,Singha4}.

\begin{figure}
\includegraphics[width=0.5\textwidth]{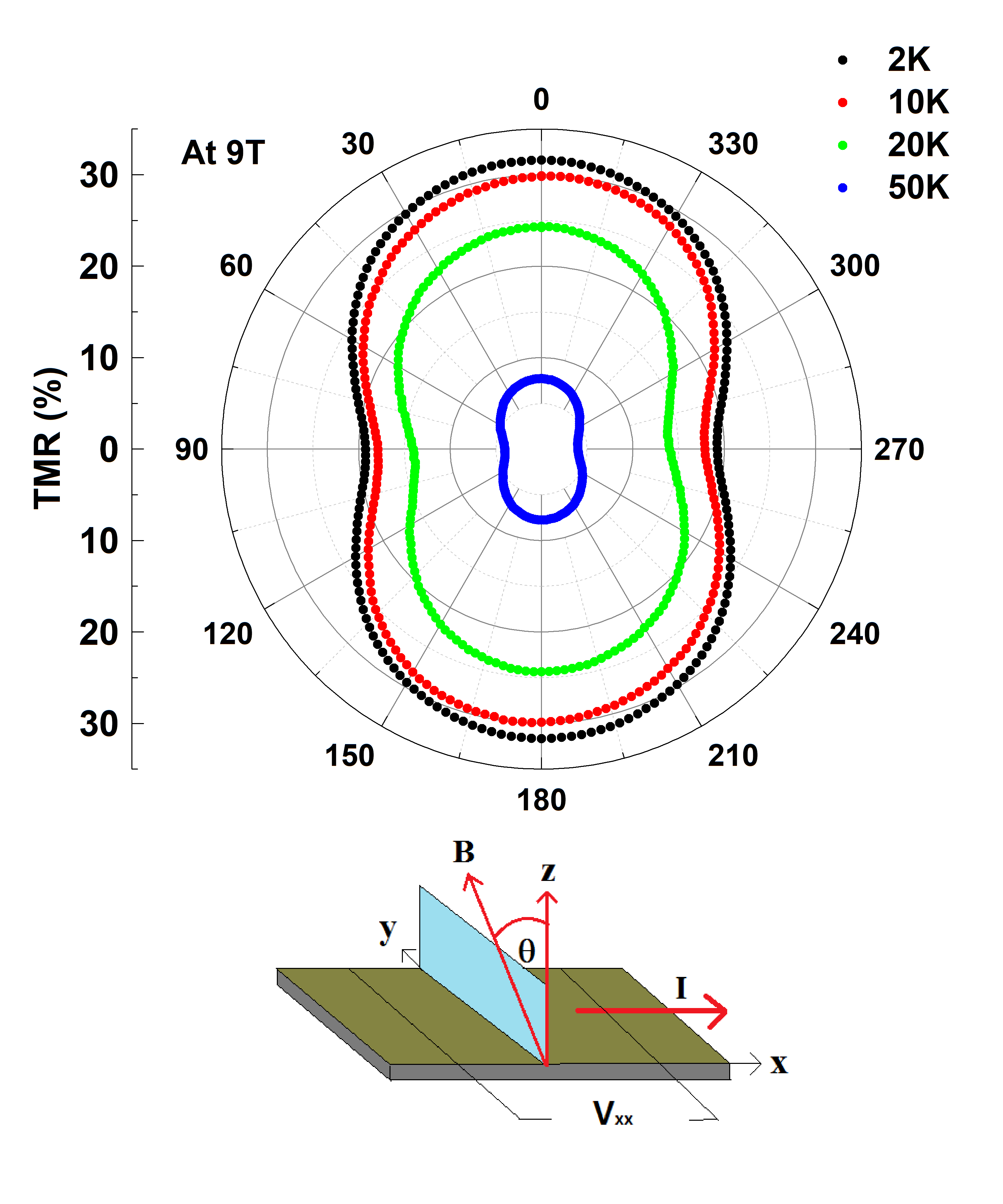}
\caption{(upper panel) Angular dependence of transverse magnetoresistance of Ta$_{3}$SiTe$_{6}$ single crystal at different representative temperatures for applied field 9 T. (lower panel) Schematic shows the experimental set-up. $\theta$ is the angle between $z$-axis and magnetic field. }\label{rh}
\end{figure}

Next, we have measured MR under transverse electric and magnetic field configuration at several fixed temperatures for understanding the exact nature of field dependence of MR. As shown in Fig. 2(b), the transverse MR is positive and shows monotonic increment with field without any sign of saturation. At 2 K and 9 T, a MR of $\sim$32\% is obtained. With the increase in temperature, however, the MR decreases rapidly. As mentioned before, the ARPES results for Ta$_{3}$SiTe$_{6}$ \cite{Sato} confirm the presence of nonsymmorphic symmetry-protected nodal-line crossing near E$_{F}$. In addition, it is shown that there is a large trivial Fermi pocket in this material. The contribution from this trivial pocket may suppress the MR in Ta$_{3}$SiTe$_{6}$ as compared to several other topological semimetals. However, the non-saturating nature of the MR in this compound still cannot be explained using semiclassical theory for trivial band structure\cite{Singha5,McKenzie}. Moreover, the observed planar Hall effect (to be discussed later on), which can only originate from relativistic chiral anomaly in a non-magnetic system, provides an indirect signature of the non-trivial band structure in Ta$_{3}$SiTe$_{6}$.

In Fig. 3, the angular dependence of transverse MR is shown for different temperatures at 9 T. In this experimental set up, the current ($I$) direction has been kept fixed along the $x$-axis within the plane of the rectangular plate-like crystal, and the magnetic field ($B$) is rotated from out-of-plane (along $z$-axis) to in-plane direction ($y$-axis), i.e., $B$ is rotated about $x$-axis [illustrated in the schematic of Fig. 3]. We observe that the MR is maximum, when the magnetic field direction is along the out-of-plane direction, i.e., ($\theta\simeq$0) and becomes minimum, when the field is along in-plane direction ($\theta\simeq$90$^{\circ}$). The polar plot shows a two-fold symmetric pattern. At 2 K and 9 T, the anisotropy ratio is about 1.5, which is quite small given the predicted quasi two-dimensional electronic structure of Ta$_{3}$SiTe$_{6}$.

\begin{figure}
\includegraphics[width=0.5\textwidth]{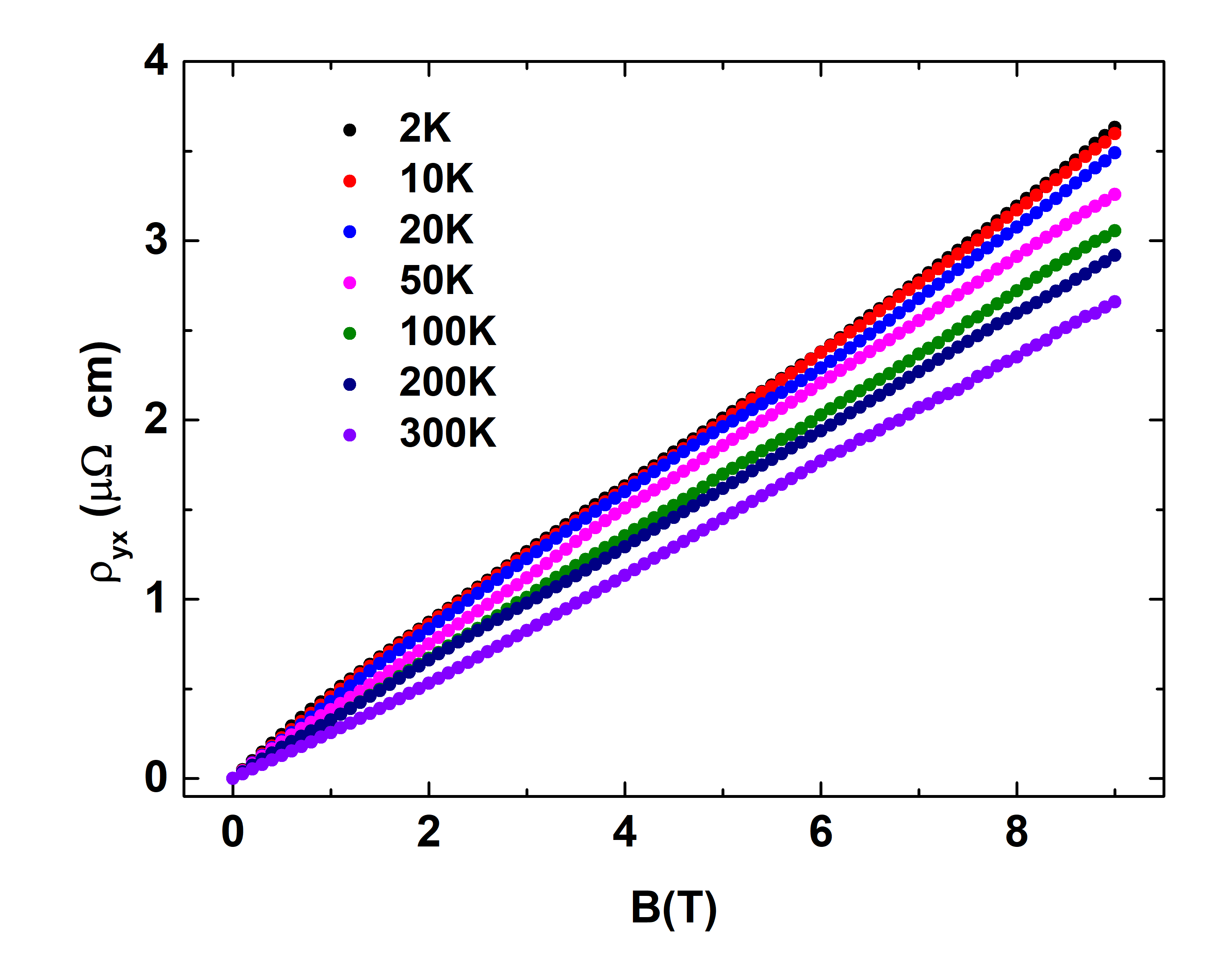}
\caption{ Hall resistivity for Ta$_{3}$SiTe$_{6}$ as a function of magnetic field at different temperatures.}\label{rh}
\end{figure}

To determine the type and density of charge carriers and their mobility, we have performed the Hall resistivity ($\rho_{yx}$) measurement as functions of temperature and magnetic field. In Fig. 4, $\rho_{yx}$ has been plotted as a function of magnetic field at some representative temperatures. At all temperatures, $\rho_{yx}$ is almost linear in field with a positive slope, which indicates hole dominated conduction. From the semiclassical one-band model, the calculated hole density ($n_{h}$) at 2 K and 9 T is $\sim$1.7$\times$10$^{21}$ cm$^{-3}$ and mobility ($\mu_{h}$) is $\sim$0.6$\times$10$^{3}$ cm$^{2}$ V$^{-1}$ s$^{-1}$. In a nodal-line semimetal, the band crossing occurs along a line in momentum space in contrast to discrete points in a Dirac/Weyl semimetal. As a result, if such a crossing resides at the Fermi energy, the corresponding Fermi pocket would be quite larger compared to a tiny pocket originated from the Dirac/Weyl point crossing. Hence, the carrier density of a nodal-line semimetal ($\sim$10$^{21}$-10$^{22}$ cm$^{-3}$) is usually quite higher than that of a Dirac/Weyl semimetal ($\sim$10$^{17}$-10$^{19}$ cm$^{-3}$).

The obtained value of carrier density in Ta$_{3}$SiTe$_{6}$ is comparable to that observed in some topological nodal-line semimetals \cite{Laha,Sankar}. The high carrier density might provide an additional support for the possible nodal-line state in Ta$_{3}$SiTe$_{6}$.

\begin{figure}
\includegraphics[width=0.5\textwidth]{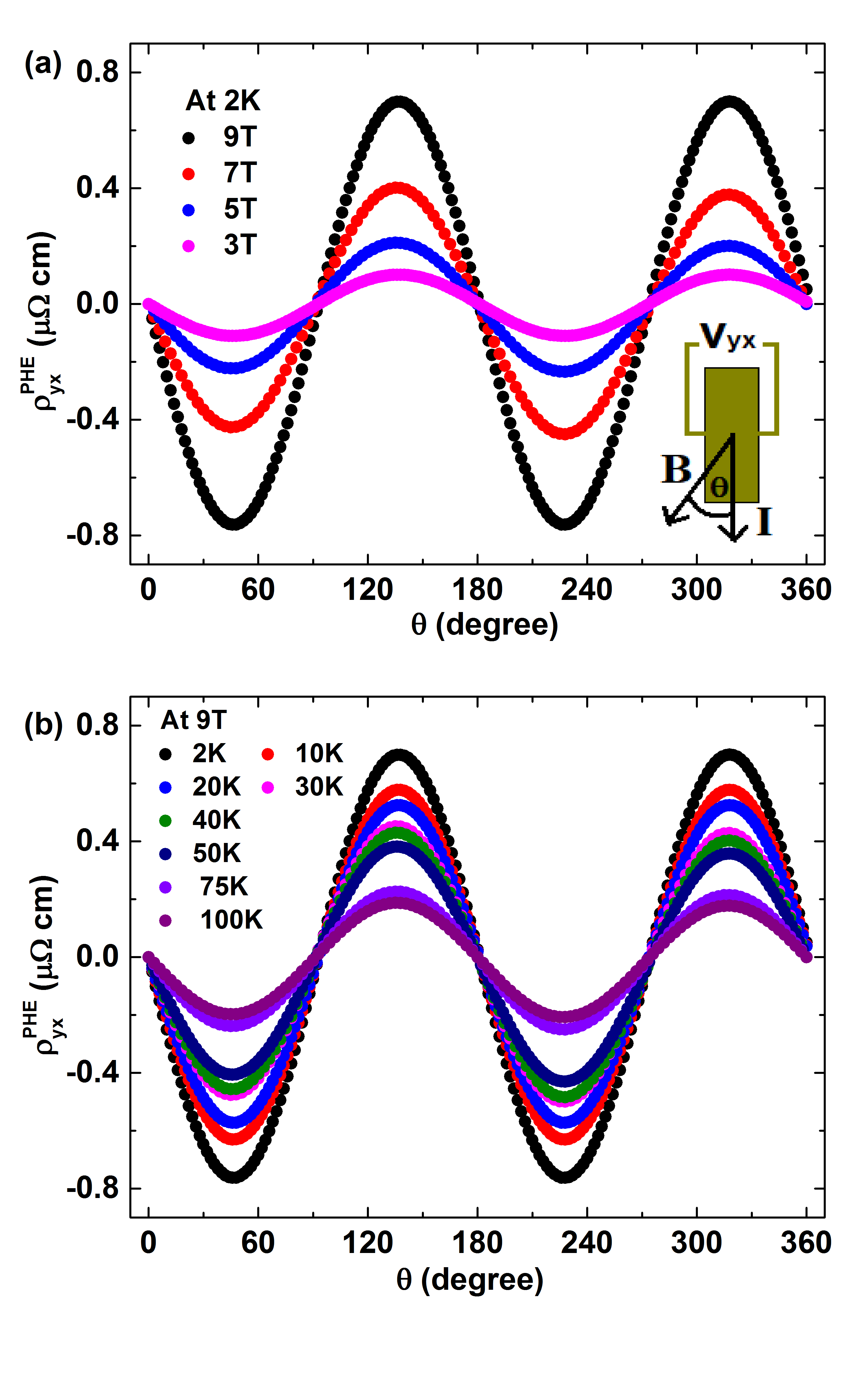}
\caption{ (a) Angle dependence of the planar Hall resistivity ($\rho_{yx}^{PHE}$) for Ta$_{3}$SiTe$_{6}$ for different magnetic fields at 2 K. Inset shows the schematic for experimental set-up. $\theta$ is the angle between current and magnetic field.(b) Angle dependence of $\rho_{yx}^{PHE}$ at different temperatures at 9 T.}\label{rh}
\end{figure}

In Fig. 5(a), we have shown the angle dependence of the planar Hall resistivity ($\rho_{yx}^{PHE}$) for Ta$_{3}$SiTe$_{6}$ at 2 K for different magnetic field strengths. In the inset, the schematic illustrates the experimental setup for PHE measurement, where the magnetic field lies on the same plane of the current and voltage ($V_{yx}$). The magnetic field is rotated within the plane. Due to the nature of Lorenz force, the conventional Hall resistivity is zero when the magnetic field and electric field (i.e. current) are coplanar and is maximum, when magnetic field, current, and Hall voltage are mutually perpendicular. The experimental configuration of PHE suggests that there should not be any Lorenz force. However, to exclude any contribution from conventional Hall voltage due to small misalignment, $\rho_{yx}^{PHE}$ measurements have been performed with both positive and negative field directions and $\rho_{yx}^{PHE}$ has been calculated from the average. As demonstrated in Fig. 5(a), $\rho_{yx}^{PHE}$ shows a periodic nature with a period of $\pi$ and the amplitude increases with the increasing field. It shows maxima at 135$^{\circ}$ and 315$^{\circ}$, whereas minima at 45$^{\circ}$ and 225$^{\circ}$. The positions of extrema are consistent with the theoretically predicted PHE \cite{Nandy}. As shown in Fig. 5(b), though $\rho_{yx}^{PHE}$ decreases with increasing temperature, the amplitude shows weak temperature dependence. $\rho_{yx}^{PHE}$ remains detectable at 100 K and above. This is one of the remarkable differences between PHE and chiral anomaly induced negative LMR as it vanishes quickly with increasing temperature \cite{Huang,Li,Li2,Pariari,Wang}.

\begin{figure}
\includegraphics[width=0.5\textwidth]{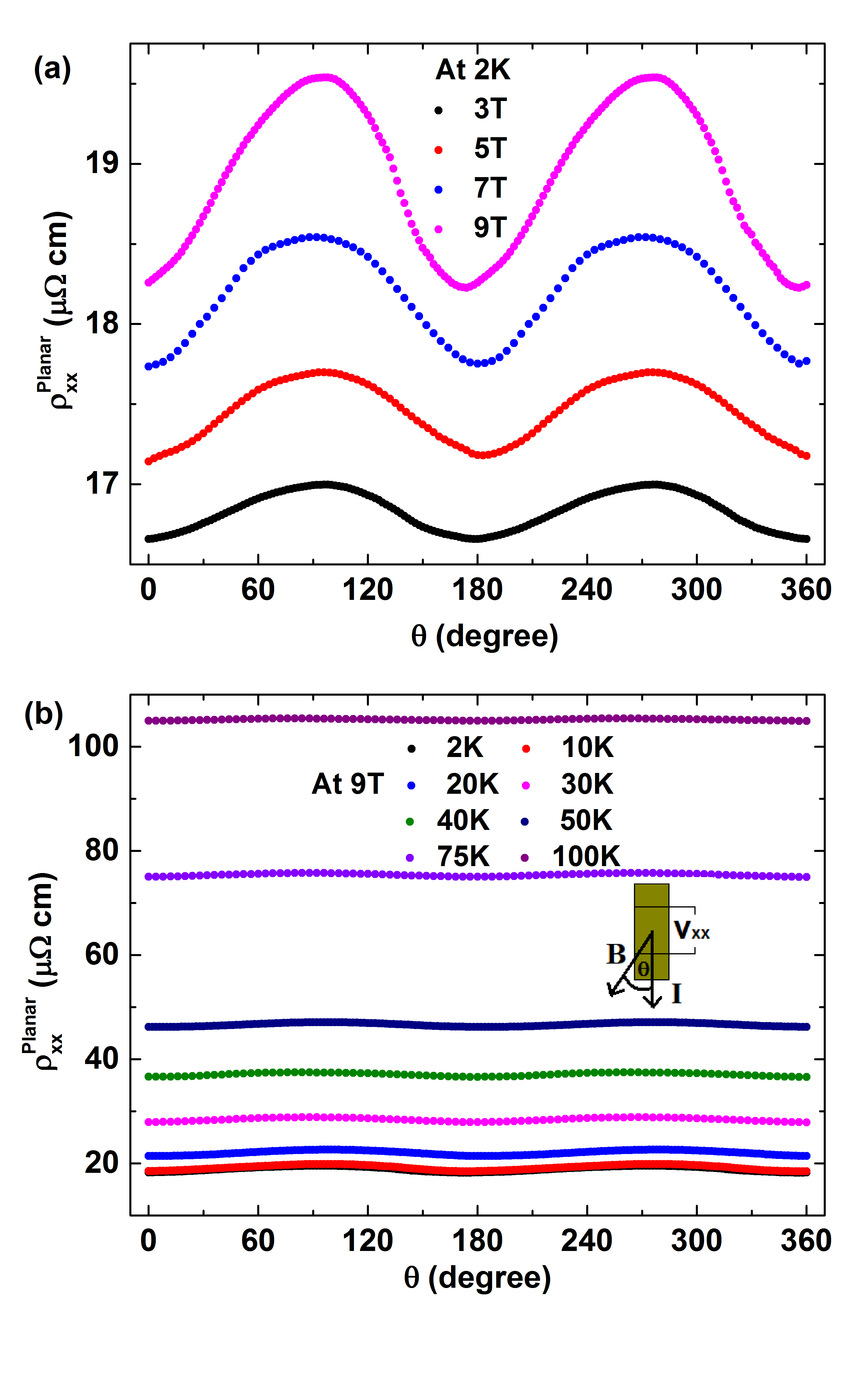}
\caption{ (a) Angle dependence of the planar resistivity ($\rho_{xx}^{planar}$) for Ta$_{3}$SiTe$_{6}$ for different magnetic fields at 2 K. (b) Angle dependence of $\rho_{xx}^{planar}$ at different temperatures at 9 T. The measurement set-up is shown schematically in the inset.}\label{rh}
\end{figure}

The planar resistivity ($\rho_{xx}^{planar}$) for different temperatures and magnetic fields have been shown in Fig. 6(a) and Fig. 6(b). The inset illustrates the experimental setup. Likewise $\rho_{yx}^{PHE}$, $\rho_{xx}^{planar}$ also shows a periodicity of $\pi$. However, the positions of extrema are different. $\rho_{xx}^{planar}$ is maximum at 90$^{\circ}$ and 270$^{\circ}$, i.e., when magnetic field and current are in transverse configuration.
 PHE in topological systems can be mathematically formulated from the semiclassical Boltzmann theory \cite{Burkov,Nandy} as
\begin{equation}
\rho_{yx}^{PHE}= -\Delta\rho_{chiral}\sin\theta\cos\theta,
\end{equation}
\begin{equation}
\rho_{xx}^{planar}= \rho_{\bot}-\Delta\rho_{chiral}\cos^{2}\theta,
\end{equation}
where  $\Delta\rho_{chiral}=\rho_{\bot}-\rho_{\|}$ is the chiral anomaly induced resistivity component. $\rho_{\bot}$ and $\rho_{\|}$ are the resistivity for transverse and longitudinal configuration, respectively. Although, this theoretical model of PHE is formulated for Dirac/Weyl semimetals, the fundamental arguments remain also valid in nodal-line semimetal state; hence we have used this model to analyze the PHE in Ta$_{3}$SiTe$_{6}$.

\begin{figure*}
\includegraphics[width=1\textwidth]{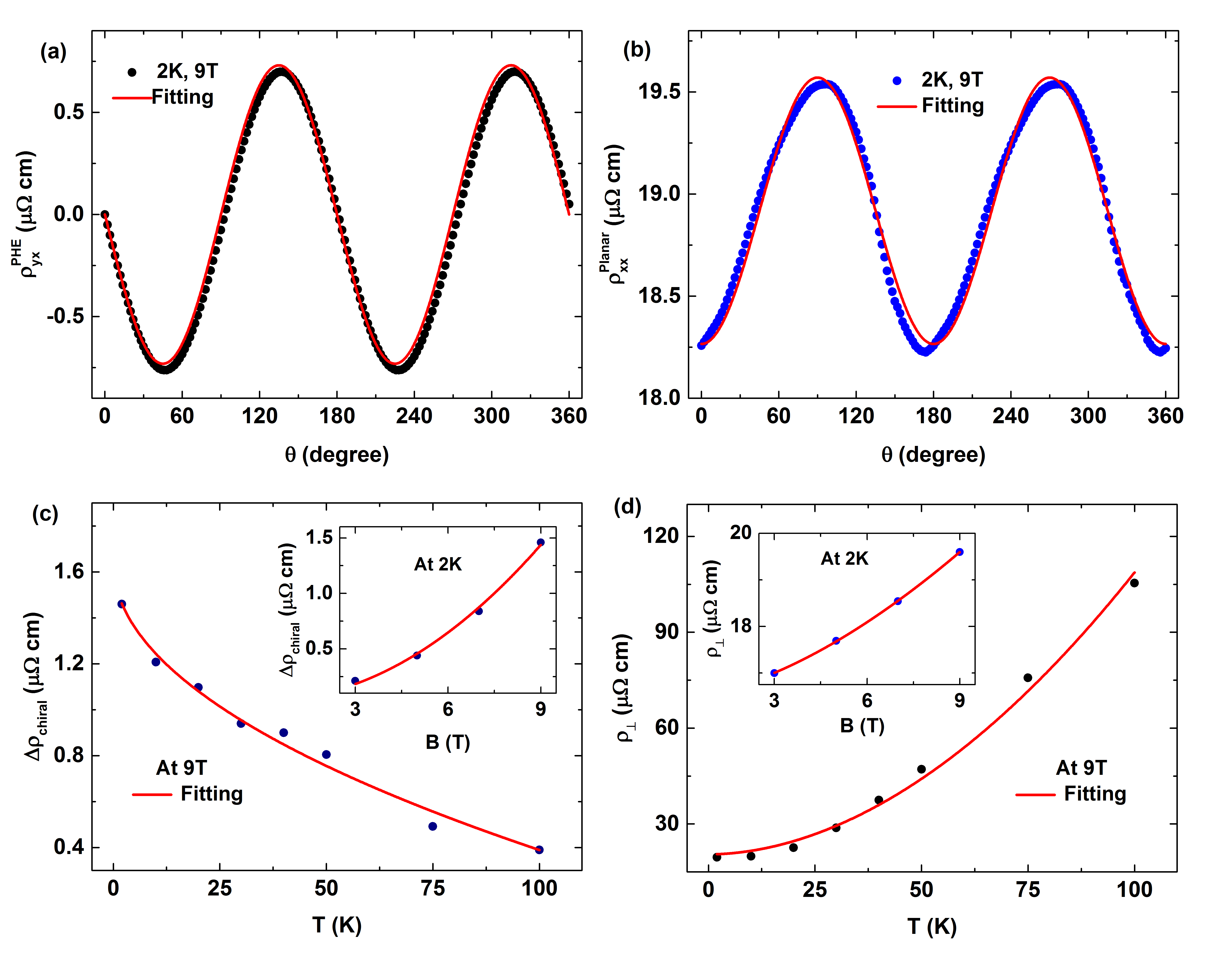}
\caption{ Fittings of (a) $\rho_{yx}^{PHE}$ and (b) $\rho_{xx}^{planar}$ for Ta$_{3}$SiTe$_{6}$ using Eqs. (1) and (2), respectively. (c) The temperature dependence of the extracted chiral anomaly induced resistivity component ($\Delta\rho_{chiral}$) at 9 T. Inset shows magnetic field dependence of $\Delta\rho_{chiral}$ at 2 K. (d) The temperature dependence of the extracted transverse resistivity ($\rho_{\bot}$) at 9 T. Inset shows magnetic field dependence of $\rho_{\bot}$ at 2 K.}\label{rh}
\end{figure*}

Using the theoretical Eqs. (1) and (2), we have fitted $\rho_{yx}^{PHE}$ and $\rho_{xx}^{planar}$ in Figs. 7(a) and 7(b), respectively. From the fitting parameters, the chiral resistivity and transverse resistivity components have been extracted. In Figs. 7(c) and 7(d), the temperature dependence of $\Delta\rho_{chiral}$ and $\rho_{\bot}$ have been plotted. At 9 T, $\Delta\rho_{chiral}$ decreases with increasing temperature, showing a $\Delta\rho_{chiral}=A_{1}-A_{2}T^{n}, (n\sim1/2)$ type power-law behavior, where $A_{1}$ and $A_{2}$ are constants, while, $\rho_{\bot}$ increases almost quadratically with temperature. Such monotonic decrease of $\Delta\rho_{chiral}$ with the increase in temperature has been observed in several Dirac/Weyl semimetals \cite{Kumar, Wang2, Singha3, Li3, Li4}. Both $\Delta\rho_{chiral}$ and $\rho_{\bot}$ have been fitted to a power-law function of magnetic field ($B$), i.e., $\Delta\rho_{chiral}\propto B^{2}$ and $\rho_{\bot} \propto B^{1.6}$ [Insets of Figs. 7(b) and 7(d)]. The deduced values of exponents are very close to those observed in other systems \cite{Singha3, Li4, Li5, Liang, Wang2}. However, theoretically it was predicted that $\Delta\rho_{chiral}$ does not follow a simple linear or quadratic field dependence rather its dependency can be divided into two different magnetic field regions \cite{Burkov}.
\begin{equation}
\Delta\rho_{chiral} \propto (\frac{L_{c}}{L_{a}})^{2} \propto B^{2},
\end{equation}
for  $L_{a}\gg L_{c}$; weak magnetic-field region.
\begin{equation}
\Delta\rho_{chiral} \propto \frac{1}{\sigma}(1-\frac{2L_{a}}{L_{x}}),
\end{equation}
for $L_{a}\ll L_{c}$, $L_{a}<L_{x}<\frac{L_{c}^{2}}{L_{a}}$; strong magnetic field region and
\begin{equation}
\Delta\rho_{chiral} \propto \frac{1}{\sigma}(1-\frac{L_{a}^{2}}{L_{c}^{2}}),
\end{equation}
for $L_{a}\ll L_{c}$, $L_{x}>\frac{L_{c}^{2}}{L_{a}}$; strong magnetic field region.\\

where, $\sigma$, $L_{a}$, $L_{c}$, $L_{x}$ are conductivity, magnetic length ($L_{a}\propto B^{-1}$), chiral charge diffusion length, and sample length, respectively \cite{Burkov}. $\Delta\rho_{chiral}$ follows a $B^{2}$ dependence in the low field region while at strong field the amplitude of PHE will follow either $B^{-1}$ [Eq. (4)] or $B^{-2}$ [Eq. (5)], depending on the sample length $L_{x}$. As we can see from Fig. 7(b)(inset) that $\Delta\rho_{chiral}$ follows almost $B^{2}$ dependence up to 9 T, which is satisfied in low field region limit. It is important to note that the quantum oscillation frequency for Ta$_{3}$SiTe$_{6}$ has been reported to be very high, $\sim$1383 T \cite{Naveed}. For such a large Fermi pocket, it is expected that one would require really high magnetic field to reach the quantum limit. Hence, the low field approximation should be valid in the present case. Our results call for further investigation in high magnetic field. A few previous reports \cite{Kumar, Liang, Li5} on PHE suggest a possible dominance of strong anisotropic orbital MR (OMR) over the chiral anomaly. It is well known that the transport properties of materials are closely related to the Fermi surface topology. If the morphology of Fermi pockets is complicated enough, transport parameters such as effective mass, mean scattering time, mobility, etc.  would also vary along different directions. The same feature is also noticed in magnetoresistance in different directions. This effect may be commonly observed in high MR along with highly anisotropic systems. However, previous reports on Ta$_{3}$SiTe$_{6}$ \cite{Naveed} as well as our present study show that the transverse MR  is not very large and highly anisotropic (e.g. TMR is $\sim$32\% and anisotropy is about 1.5 for transverse MR, whereas for planar MR it is about 1.06) compared to the systems, where such OMR dominant PHE was reported \cite{Kumar, Liang, Li5}.

\section{Conclusion}

In conclusion, we have performed magnetotransport measurements on single crystals of Ta$_{3}$SiTe$_{6}$. A nonsaturating magnetoresistance has been observed. The linear field dependence of Hall resistivity shows the presence of hole type charge carrier in the system. In spite of having comparatively high carrier density, the mobility of charge carriers is moderately high. In our measurements, prominent planar Hall effect has been observed, which is robust and persists up to high temperature. Planar Hall originates from the relativistic Adler-Bell-Jackiw chiral anomaly and nontrivial Berry curvature. Therefore, these results support the nontrivial nature of the band structure in the material and validate the presence of Dirac-like dispersions. As the bulk materials of layered ternary telluride $X_{3}$SiTe$_{6}$ ($X$=Ta, Nb) have been predicted to host accidental Dirac-loops and essential fourfold nodal-lines in the absence of SOC and fourfold degenerate hourglass Dirac-loop in presence of SOC, a family of isostructural materials with identical electronic band structures provides a great advantage for tunability of their electronic states. For example, by substitution of atoms of different atomic numbers ($Z$) from Nb ($Z$=41) to Ta ($Z$=73), one can tune SOC in these materials, hence their electronic properties. Likewise, by doping a small amount holes or electrons, one can move the nodal-line or the necks of the hourglass dispersions nearer to Fermi energy, resulting more exotic transport properties. So, our investigations not only ratify the presence of topological nontrivial band structure in Ta$_{3}$SiTe$_{6}$, but also offer an excellent platform for further investigations by chemical substitution/doping.

\section{ACKNOWLEDGMENTS}

We thank Dr. Dipten Bhattacharya for his help in EDAX analysis. We also thank Arun Kumar Paul for his help during sample preparation and measurements.

\end{document}